# Joint Design of Congestion Control Routing With Distributed Multi Channel Assignment in Wireless Mesh Networks

K.Valarmathi,

Research Scolar, Sathyabama University,
Chennai, India

valarmathiphd@gmail.com

Dr. N.Malmurugan

Principal, Oxford Engineering Collge,
Trichy, India

*Abstract*— **In Wireless Mesh Networks (WMN), a channel assignment has to balance the objectives of maintaining connectivity and increasing the aggregate bandwidth. The main aim of the channel assignment algorithm is to assign the channels to the network interfaces, from the given expected load on each virtual link. From the existing work done so far, we can examine that there is no combined solution of multi-channel assignment with routing and congestion control. In this paper, we propose a congestion control routing protocol along with multi-channel assignment. We use a traffic aware metric in this protocol in order to provide quality of service. The proposed protocol can improve the throughput and channel utilization to very high extent because it provides solution for multi-channel assignment and congestion control. The proposed algorithm assigns the channels in a way that, congestion is avoided and co-channel interference levels among links with same channel are reduced. By our simulation results in NS2, we show that the proposed protocol attains high throughput and channel utilization along with reduced latency.**

## I. INTRODUCTION

Wireless mesh networks (WMNs) contains many stationary wireless routers which are interconnected by the wireless links. For the wireless mobile devices these wireless routers acts as the access points (APs) and some of them act as gateways to the internet through the high speed wired links. Wireless mobile devices transfer data to the associated wireless router and these data are then transferred in a multi-hop manner to the internet through the intermediate wireless routers. Wireless mesh networks are popular because they are capable to credit their low cost and their auto organizing features [1].

Topology discovery, traffic profiling, channel assignment and routing are required in the multi-channel wireless mesh network architecture [3]. The static aggregation nodes which are similar to wireless LAN access points are comprised in the multi-channel wireless mesh networks. Since the 802.11b interface hardware handles the bandwidth problem efficiently, it is used in the construction of the multi-channel mesh network. Each node in a multi-channel wireless mesh network consists of multiple 802.11 complaint NICs. Each of these is tuned to a particular radio channel for a long period of time such as hours or days [3].

The network congestion is caused due to the interference between neighboring links. The aggregated throughput will be reduced for the TCP data connections if the links becomes congested. To reduce the interference due to the neighboring transmission of the same channel, an efficient channel assignment algorithm is necessary [2].

Maintaining connectivity and increasing aggregate bandwidth are the two goals, a channel assignment algorithm has to balance. Binding each network interface to a radio channel is the objective of the channel assignment in a multi-channel wireless mesh network. This is done in a way such that the available bandwidth on each virtual link is proportional to its expected load. The channel assignment is based on the network topology information [3]. With the help of some outside agency and by the changes which occurs rarely, the channel assignment is estimated [4]. To support the dynamic channel assignment, the fixed channel assignment scheme can be easily extended [5].

The routing consists of traffic from the mobile nodes to any access point (called Internet gateway) and response traffic back to the mobile nodes [7]. Routing is based on the scalar potential fields and not on the route entries. Mostly alternative routes are estimated without additional overhead when the primary route fails. The routing algorithms mainly focus on improving the network capacity or the performance of individual transfers rather than copying with mobility or reducing the power utility [8].

From the existing work done so far, we can observe that, no work is done on the combined solution of multi- channel assignment with routing protocol and congestion control. Also the use of a traffic aware metric can be investigated which could optimize the capacity of the network.

In this paper, we propose a congestion control routing protocol along with multi-channel assignment. In this protocol, we use traffic aware metric to provide quality of service. Since it provides solution for multi channel assignment and congestion control, the proposed protocol can improve the throughput and channel utilization to very high extent.

.

## II. RELATED WORK

The Joint Optimal Channel Assignment and Congestion Control (JOCAC) [2] is formulated as a decentralized utility maximization problem with constraints that arise from the interference of the neighboring transmissions. The focus is on improving the aggregate capacity of the IEEE 802.11a/b/g-based WMNs via the use of multiple channels in each router. The JOCAC algorithm allocates channels to control the interference on each link regarding the link's average congestion price. One of the distinct advantages of this algorithm is the ability to assign not only the non-overlapping channels, but also the partially overlapping channels.

Richard Draves et al [4], have proposed a new metric for routing in multi-radio, multihop wireless networks is proposed. The goal of the metric is to choose a high-throughput path between a source and a destination. Our metric assigns weights to individual links based on the Expected Transmission Time (ETT) of a packet over the link. The ETT is a function of the loss rate and the bandwidth of the link. The individual link weights are combined into a path metric called Weighted Cumulative ETT (WCETT) that explicitly accounts for the interference among links that use the same channel. The WCETT metric is incorporated into a routing protocol that we call Multi-Radio Link-Quality Source Routing. A routing protocol MR-LQSR (Multi-Radio Link-Quality Source Routing) with a new metric WCETT (Weighted Cumulative Expected Transmission Time) is implemented to accomplish this task, and compared its performance to other routing metrics WCETT allows us to trade off channel diversity and path length

Bouckaert et al [5], have proposed an adaptive block addressing (ABA) scheme is first introduced for logic address assignment as well as network auto-configuration purpose. The scheme takes into account the actual network topology and thus is fully topology-adaptive. Then a distributed link state (DLS) scheme is further proposed and put on top of the block addressing scheme to improve the quality of routes, in terms of hop count or other routing cost metrics used, robustness, and load balancing. An efficient scalable wireless mesh routing protocol, called topology-guided distributed link state (TDLS), provides multiple paths that preclude the need for explicit route repair.TDLS comprises two basic schemes, namely, the adaptive block addressing (ABA) scheme and the distributed link state (DLS) scheme.

Baumann et al [7], have proposed HEAT, a cast routing protocol is designed to scale to the network size and to be robust to node mobility. HEAT relies on a temperature field to route data packets towards the Internet gateways, as follows: Every node is assigned a temperature value, and packets are routed along increasing temperature values until they reach any of the Internet gateways, which are modeled as heat sources. The distinguishing feature of the protocol is that it does not require flooding of control messages rather every node in the network determines its temperature considering only the temperature of its direct neighbors. The realistic mobility patterns extracted from geographical data of large Swiss cities is used.

Sumit Rangwala et al [9], have explored the mechanisms for achieving fair and efficient congestion control for multi-hop wireless mesh networks. First, they have designed an AIMD-based rate-control protocol called Wireless Control Protocol (WCP) which recognizes that wireless congestion is a neighborhood phenomenon, not a node-local one, and appropriately reacts to such congestion. Second, they have designed a distributed rate controller that estimates the available capacity within each neighborhood, and divides this capacity to contending flows, a scheme we call Wireless Control Protocol with Capacity estimation (WCPCap). Using analysis, simulations, and real deployments, they found that their designs produce rates that are both fair and efficient, and achieve near optimal goodputs for all the topologies.

Anastasios Giannoulis et al [10] have addressed the problem by introducing a formulation that allows its decomposition in two sub problems: A congestion control sub problem for traffic allocation to a fixed channel assignment over a node path and a discrete combinatorial channel assignment sub problem. They solved the conditional congestion control sub problem by mapping it to an optimization problem of traffic distribution to a set of radio paths. The solution provides channel congestion information that is utilized to address the channel assignment sub problem. This leads to an iterative procedure which guarantees successive increases to overall network utilization**.**

Angela Feistel et al [11] have been deal with the problem of joint hop-by hop congestion control and power control in wireless networks. They have coupled the back-pressure policy and window-based congestion control at source nodes with power control to provide end to- end fairness. Their definition of fairness includes arbitrarily close approximation of max-min fairness. Numerical experiments indicate good transient response to channel variations.

Tae-Suk Kim et al [12] have developed a framework to address the problem of maximizing the aggregate utility of traffic flows in wireless mesh networks, with constraints imposed both due to self interference and minimum rate requirements. The output of their framework is a schedule that dictates which links are to be activated simultaneously, and provides specifications of the resources associated with each of those links. Utilizing the proposed framework as a basis, they have build an admission control module that intelligently manages the resources among the flows in the network and admits as many new flows as possible without violating the QoS of the existing flows. They have also provided numerical results to demonstrate the efficacy of their framework.

Asad Amir Pirzada et al [13] have proposed a congestion aware routing protocol, which can successfully establish channel diverse routes through least congested areas of a hybrid WMN. The prime advantage of their protocol is its ability to discover optimal routes in a distributed manner without the requirement of an omniscient network entity. Simulation results show that the congestion aware routing protocol can successfully achieve a high packet delivery ratio with lower routing overhead and latency in a hybrid WMN.

### III. PROPOSED JOINT DESIGN PROTOCOL

The proposed algorithm allocates channels in a way that

(a) *congestion* is avoided and

(b) *co-channel interference* levels among links with same channel are reduced.

Links with higher costs are allocated to higher priorities in case of channel assignment over the links with lower cost, in our algorithm. Scheduling the links becomes harder because the links with higher costs suffer from the higher levels of congestion. Thus the proposed channel assignment algorithm begins by sorting links in the descending order of their link costs. Then the channels are assigned to the links respectively.

The channel which is assigned to a link is selected based on the sum of link gains between all the interfering senders using the same channel and the receiver of the link to improve the effects of co-channel interference. Then this sum is calculated for each of the channels and the channel which has the least associated value is selected for the link.

The main design issues involved in the placement of wireless mesh network nodes and a traffic profile which contains the traffic load between each pair of the nodes, are

(1) how to assign a radio channel to each 802.11 interface, and

(2) how to route traffic between all pairs of nodes, in such a way that the total good put of the wireless mesh network is maximized.

#### A. Routing Problem

Congestion Control rely on the expected load on each virtual link and this load is depended on the routing. The routing algorithm determines the route through the network for each communicating node pair with the given set of communicating node pairs, the expected pair, the expected traffic between them and the virtual capacities. The resulting routes govern the path taken by future traffic because it occupies the routing tables of all the nodes. Routing also plays a vital role in the load balancing of the network apart from determining the traffic route for each communicating node pair. Load balancing helps to avoid bottleneck creation in the network which increases the efficiency of the network resource consumption.

#### B. Evaluation Metric

The ultimate goal of traffic engineering a backbone network is to maximize its overall good put, or the number of bytes it can transport between nodes within a unit time. This enables the network to support more end-user flows, and in turn more number of users. To formalize this goal, we use the idea of cross-section good put of the network. The goodput G of a network is defined as

$$G = \sum_{s,d} B(s,d) \qquad (1)$$

Here, $B(s,d)$ is the useful network bandwidth assigned between a pair of ingress-egress nodes $(s,d)$. If the traffic profile has an expected traffic load of L(s, d) between the node pair $(s,d)$, then only up to $L(s,d)$ of the assigned bandwidth between the node pair $(s,d)$ is considered useful. This criterion ensures that we only count the usable bandwidth of the network towards its throughput, hence the term good put. The goal of the channel assignment and routing algorithms is to maximize this good put G.

### IV. LOAD AWARE CHANNEL ASSIGNMENT/ROUTING

#### A. Initial Link Load Estimation

The combined channel assignment and routing algorithm first derives a rough estimate of the expected link load. We assume the capacity of link $l$ $L_c$ as,

$$Lc = \frac{N_{ch} * CH_c}{n_1} \qquad (2)$$

Where $N_{ch}$ is the number of available channels, $CH_c$ is the capacity per channel and $n_l$ are the number of virtual links within the interference range $l$. The equation essentially divides the aggregated channel capacities among all interfering links, without regard to number of NICs per node. Based on these virtual link capacities, the routing algorithm determines the initial routes and thus the initial link loads.

#### B. Expected Link Load Estimation

A more accurate estimate of expected link load is based on the notion of link quality. To compute initial expected link loads, we assume perfect load balancing across all acceptable paths between each communicating node pair. Let the number of acceptable paths between a pair of nodes $(s,d)$ is $P_n$ and the number of acceptable paths between $(s,d)$ that pass a link $l$ is $P_n(l)$. Then the expected load on link $l$, $\delta_l$, is calculated using the equation

$$\delta_l = \sum_n (P_n(1) / P_n) * W \qquad (3)$$

Where W is the estimated load between the node pair (s, d) in the traffic profile. This equation says that the initial expected load on a link is the sum of loads from all acceptable paths, across all possible node pairs that pass through the link. Because of the assumption of uniform multi-path routing, the load that an acceptable path between a nodes pair is expected to carry is the node pair's expected load divided by the total number of acceptable paths between them.

#### C. Congestion Control

Let $\delta_{l1}, \delta_{l2}, \delta_{l3}, \cdots$ be the estimated loads of the links l1, l2, l3,....

Then the link cost function $C(1_i)$ is then defined as

$$C(1_i) = \infty, \quad if\ \delta_{li}, > L_{thr}$$
$$= 1 + \delta_i, if\ \delta_{li} > 0\ and\ \delta_{li} <= L_{thr}$$
$$= 1 - \delta_{li}, if\ \delta_{li} = 0 \qquad (4)$$

where $L_{thr}$ is the load threshold.

### D. Channel Assignment

The proposed algorithm allocates channels in a way that (a) self-interference is avoided and (b) co-channel interference levels among links that use the same channel are kept as low as possible. With our algorithm, links with higher costs are assigned higher priorities in terms of channel assignment over the links with lower cost. This is because links with higher costs suffer from higher levels of congestion and thus, scheduling these links is harder. The proposed channel assignment algorithm starts by sorting links in the descending order of their link costs. Then, channels are assigned to the links in that order. The proposed algorithm avoids self interference by not assigning a channel to any link whose incident links have already been assigned channels. In other words, a link is eligible for activation only if it has no active neighbor links. In order to alleviate the effects of co-channel interference, the channel that is assigned to a link is selected based on the sum of link gains between all the interfering senders using the same channel and the receiver of the link. This sum is calculated for each of the channels and the channel with the least associated value is selected for the link. The proposed channel assignment is summarized in Algorithm, where we define $Q(c)$ to be the set of links that are assigned channel c. An active link is then assigned a transmit power based on our power assignment algorithm discussed next.

In this algorithm, we can define

- $V_{(1,c)}$ as a channel assignment vector $V$ with elements $V(1,c)$.

- Let $S_1(c)$ be the set of links that are assigned channel c.

- Let $n_1(j)$ be the set of neighboring links which share either the sender or the receiver of link l.

- $LG1_j$ is the link gain between the receiver and the sender.

**Algorithm**

1: Initialization: $V_{(1,c)} \leftarrow 0, and\ S_1(c) \leftarrow 0, \forall 1 \in L$ and $\forall\ c \in C$ ;

2: Sort links by descending order of $\delta$ , and label i-th link in the sorted link as $1_i$ ;

3: $for\ j = 1\ to\ L\ do$

4: $if\ \Sigma_e \Sigma_c V(e,c) = 0, for\ e \in n_1(j)$ then

5: Calculate $d_c = \Sigma_q \in S1(c)LG1_i, \forall c \in C$;

6: Allocate channel
$$c_{1i} = \min_c \{d_1, d_2 \cdots d_c\} to\ link\ 1_i\ ;$$

7: $Assign 1_i\ to\ S_1(c_{1i})$ ;

8: $end\ if$

9: $end\ for$

## V. Simulation Results

### A. Simulation Model and Parameters

We use NS2 [14] to simulate our proposed protocol. We use the IEEE802.16e simulator [15] patch for NS2 version 2.33 to simulate a WiMax Mesh Network. It has the facility to include multiple channels and radios. It supports different types of topologies such as chain, ring, multi ring, grid, binary tree, star, hexagon and triangular. The supported traffic types are CBR, VoIP, Video-on-Demand (VoD) and FTP. In our simulation, 4 mobile nodes are arranged in a ring topology of size 500 meter x 500 meter region. All nodes have the same transmission range of 250 meters. In our simulation, the speed is set as 5m/s. A total of 3 traffic flows (two VoIP and one VoD) are used.

Our simulation settings and parameters are summarized in table 1.

TABLE I. SIMULATION SETTINGS

| No. of Nodes | 50 |
|---|---|
| Area Size | 1500 X 300 |
| Mac | 802.16e |
| Radio Range | 250m |
| Simulation Time | 100 sec |
| Traffic Source | VoIP and VoD |
| VoD Packet Size | 65536 |
| VoD Rate | 150Kb |
| VoIP Codec | GSM.AMR |
| No. of VoIP frames per packet | 2 |
| Topology Type | Ring |
| OFDM Bandwidth | 10 MHz |

### B. Performance Metrics

We compare our CCMCA protocol with the JOCAC [2] protocol. We evaluate mainly the performance according to the following metrics, by varying the simulation time and the number of channels.

**Average end-to-end delay:** The end-to-end-delay is averaged over all surviving data packets from the sources to the destinations.

**Average Packet Delivery Ratio:** It is the ratio of the number of packets received successfully and the total number of packets sent

**Throughput:** It is the number of packets received successfully.

## A. Effect of Varying Channels

In our initial experiment, we vary the number of channels as 1,2,3,4 and 5

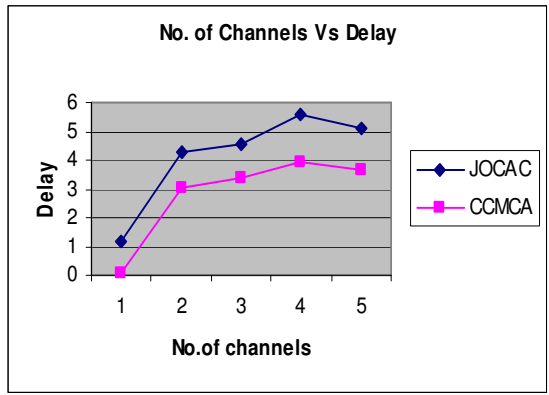

Figure 1. No. of channels Vs Delay

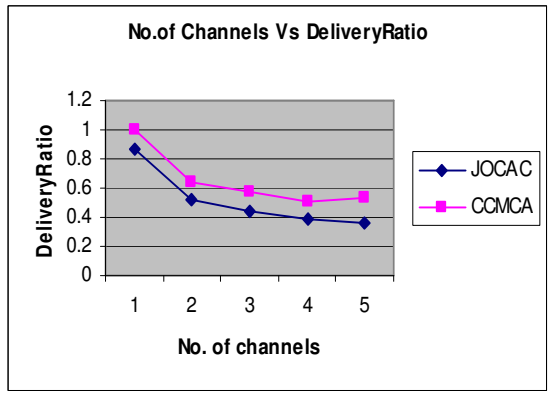

Figure 2. No. of channels Vs Delivery Ratio

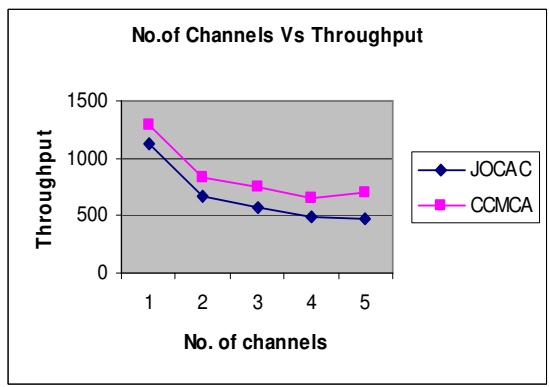

Figure 3. No. of channels Vs Delay

From Figure 1, when the number of channels increases, the average end-to-end delay also increases. We can see that the average end-to-end delay of the proposed CCMCA protocol is less when compared to the JOCAC protocol.

Figure 2 presents the packet delivery ratio of both the protocols. When the number of channels increases the packet delivery ratio decreases. We can observe that CCMCA achieves good delivery ratio, when compared to JOCAC.

Figure 3 gives the throughput of both the protocols when the number of channels is increased. When the number of channels increases the throughput decreases. As we can see from the figure, the throughput is more in the case of CCMCA, than JOCAC.

## B. Effect of varying time

In our second experiment, we vary the simulation time as 5,10,15,20 and 25.

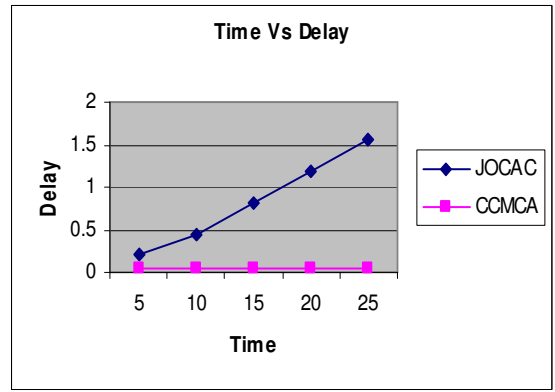

Figure 4. Time Vs Delay

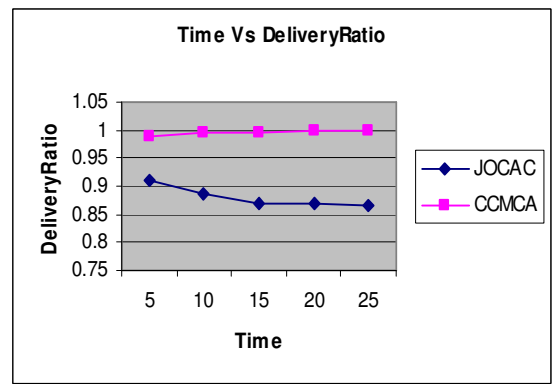

Figure 5. Time Vs Delivery Ratio

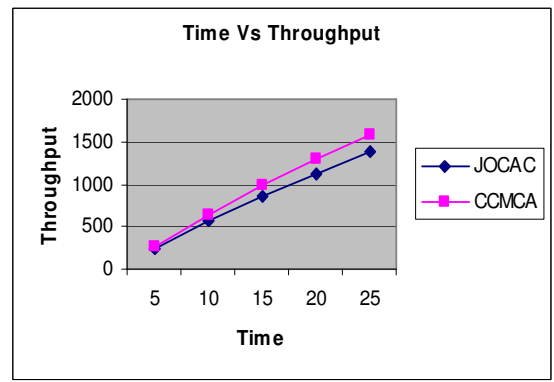

Figure 6. Time Vs Throughput

From Figure 4, when the simulation time increases, the average end-to-end delay increases for JOCAC and remains constant for CCMCA. Also we can see that the average end-to-end delay of the CCMCA protocol is less when compared to the JOCAC protocol.

Figure 5 presents the packet delivery ratio of both the protocols. When simulation time increases the packet delivery ratio gradually increases for CCMCA and decreases for JOCAC. Also we can observe that CCMCA achieves good delivery ratio, when compared to JOCAC.

Figure 6 gives the throughput of both the protocols when the number of channels is increased. When the simulation time increases then the throughput of both the protocols also increases. As we can see from the figure, the throughput is more in the case of CCMCA, than JOCAC.

## VI. CONCLUSION

In Wireless Mesh Networks (WMN), a channel assignment has to balance the objectives of maintaining connectivity and increasing the aggregate bandwidth. In this paper, we have proposed a congestion control routing protocol along with multi-channel assignment. We have used a traffic aware metric in this protocol in order to provide quality of service. Our proposed protocol improves the throughput and channel utilization to very high extent since it provides solution for multi-channel assignment and congestion control. Our proposed algorithm assigns the channels such that congestion is avoided and co-channel interference levels among links with same channel are reduced. We have presented a load-aware channel assignment algorithm where the virtual links are arrived in the wireless mesh networks in the decreasing order of the link criticality or the expected load on a link. By our simulation results in NS2, we have shown that our proposed protocol attains high throughput and channel utilization along with reduced latency. As a future work, we wish to minimize the power consumption and to achieve fairness for the multimedia flows.

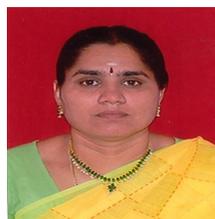

**K. Valarmathi** received BSC (Computer Science), MCA degrees from University of Madras in the year 1991, 1998 respectively. She has also received ME (Computer Science) degree in 2004 from Sathyabama University. She has worked in several Public Limited companies for more than 7 years. She worked as a Lecturer in MNM Jain Engineering College for more than 2 years. Now she is working as Assistant Professor in Panimalar Engineering College from July 2003 onwards. During her teaching service she has received Best Teacher award in the year 2007 and received cash awards many times for her academic excellence. She is currently pursuing the PhD degree in the field of Wireless Mesh Networks under the guidance of Dr. N. Malmurugan, Principal of Oxford Engineering College, Trichy.

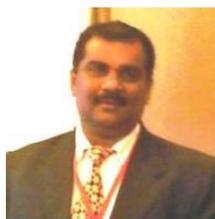

**Dr. Malmurugan Nagarajan** has around Twenty two years of both Teaching and Industrial experience. He worked extensively in Audio, Video Codec development for mobile platforms and well versed with Audio and Video standards. He is strong in developing tools & system applications and Modeling & Simulation. Currently he is developing new algorithms and IPs for the components in Wireless Broadband Physical Layer He is an Educationist and held various positions ranging from Lecturer to Principal in various educational organizations. His area of interest includes Wavelet variants based Signal and Image Processing, Multimedia Compression & Watermarking, Signal Processing algorithm development in Telecom domain. He is Member of IEEE, Fellow of IETE, Fellow of UWA and Editor of Journal of Simulation and Modeling.